\documentclass[prb,twocolumn,showpacs]{revtex4}
\pdfoutput=1
\usepackage{amssymb}
\usepackage{graphicx,amsmath}

\usepackage{bm,color,mathrsfs,hyperref}

\hypersetup{
    colorlinks=true, 
    linkcolor=blue,  
    citecolor=blue,  
}

\newcommand{\bea}{\begin{eqnarray}}
\newcommand{\eea}{\end{eqnarray}}
\newcommand{\beq}{\begin{equation}}
\newcommand{\eeq}{\end{equation}}
\newcommand{\benu}{\begin{enumerate}}
\newcommand{\enu}{\end{enumerate}}

\newcommand{\Om}{\Omega}

\newcommand{\dl}{\delta}
\newcommand{\lam}{\lambda}

\newcommand{\ptl}{\partial}

\newcommand{\bq}{{\bf q}}
\newcommand{\bM}{{\bf M}}
\newcommand{\br}{{\bf r}}
\newcommand{\bL}{{\bf L}}
\newcommand{\bu}{{\bf u}}

\begin{document}
\title{Magnetoelastic Effects in Iron Telluride}

\date{\today}
\author{I. Paul$^1$, A. Cano$^2$ and K. Sengupta$^3$}
\affiliation{
$^1$Institut N\'{e}el, CNRS/UJF, 25 avenue des Martyrs, BP 166, 38042 Grenoble, France. \\
$^2$European Synchrotron Radiation Facility, 6 rue Jules Horowitz, BP 220, 38043 Grenoble, France.\\
$^3$Theoretical Physics Department, Indian Association for the Cultivation of Science,
Kolkata-700032, India.
}

\begin{abstract}
Iron telluride doped lightly with selenium is known to undergo a first order magneto-structural
transition before turning superconducting at higher doping.
We study the effects of magneto-elastic couplings on this transition using symmetry
considerations. We find that the magnetic order parameters are coupled to the uniform monoclinic
strain of the unit cell with one iron per cell, as well as to the phonons at high symmetry points
of the Brillouin zone. In the magnetic phase the former gives rise to monoclinic distortion
while the latter induces dimerization of the ferromagnetic iron chains due to alternate lengthening
and shortening of the nearest-neighbour iron-iron bonds. We compare this system with the iron
arsenides and propose a microscopic magneto-elastic Hamiltonian which is relevant for all the
iron based superconductors.
We argue that this describes electron-lattice coupling in a system where
electron-electron interaction is crucial.

\end{abstract}

\maketitle

\section{Introduction}
\label{sec:intro}

Selenium-doped iron telluride, FeTe$_{1-x}$Se$_x$, is one of the several classes of iron based materials
which are currently being studied for their intriguing superconducting
properties.~\cite{hsu-fc,yeh-kw,sales,mizuguchi,klein}
This material, the so-called 11 system, exhibits a magneto-structural transition at low carrier
doping that is suppressed by the emergence of superconductivity at large
enough doping (or under applied pressure).~\cite{garbarino}
This feature of the phase diagram is shared by the rest of iron-based superconductors,~\cite{review}
namely the 1111 systems with chemical composition $Re$OFeAs where $Re$ refers to a rare earth
metal (e.g., La, Ce and Sm),~\cite{kamihara}
the 111 systems with composition $A$FeAs where $A$ refers to an alkali
metal (e.g., Na),~\cite{li-cruz}
and the 122 systems composed of $Ae$Fe$_2$As$_2$ where $Ae$ is an alkali
earth metal (e.g., Ba, Sr and Ca).~\cite{rotter}
LiFeAs is the only exception to this rule, as the parent compound is already a superconductor.~\cite{tapp}
In Ref. \onlinecite{cano} a general phase diagram for the
1111, 122 and 111 systems was proposed on the basis of symmetry considerations
which explained the salient features of the magnetic
and structural transitions.
The key ingredient for this general phase diagram is the magneto-elastic coupling between
the magnetic and the structural order parameters.
In this paper we show that similar symmetry-allowed couplings are also present in the 11 material,
and are important in the sense that they give rise to experimentally observable structural
distortions in the magnetic phase. Our description is based on a Ginzburg-Landau mean field
analysis of the magneto-structural transition of iron telluride followed by a comparison
with the iron arsenide systems. Furthermore, we propose a microscopic Hamiltonian that provides
a unified description of the magneto-elastic properties of all the iron-based superconductors,
and we discuss its implications.
\begin{figure*}[t]
\includegraphics[width=.75\textwidth]{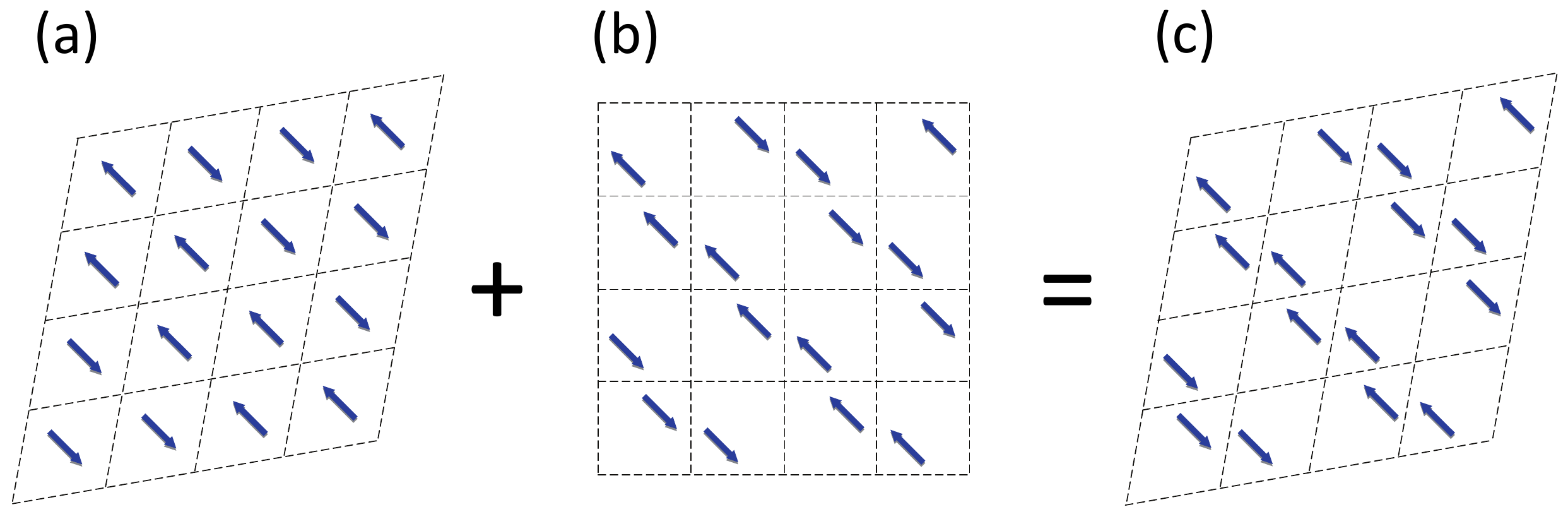}
\caption{Low temperature magneto-structural pattern on the $ab$ plane formed
by the Fe atoms of Fe$_{1+y}$Te$_{1-x}$Se$_x$. The spins order with wavevector
$(\pi/2, \pi,2)$ forming a bicollinear antiferromagnet. The associated structural
change (c) is a combination of a monoclinic distortion (a) of the square lattice,
and a dimerization (b) of the ferromagnetic Fe chains due to alternate lengthening
and shortening of the nearest-neighbour Fe-Fe bonds.}
\label{fig:1}
\end{figure*}

\begin{figure}[b]
\includegraphics[width=.4\textwidth]{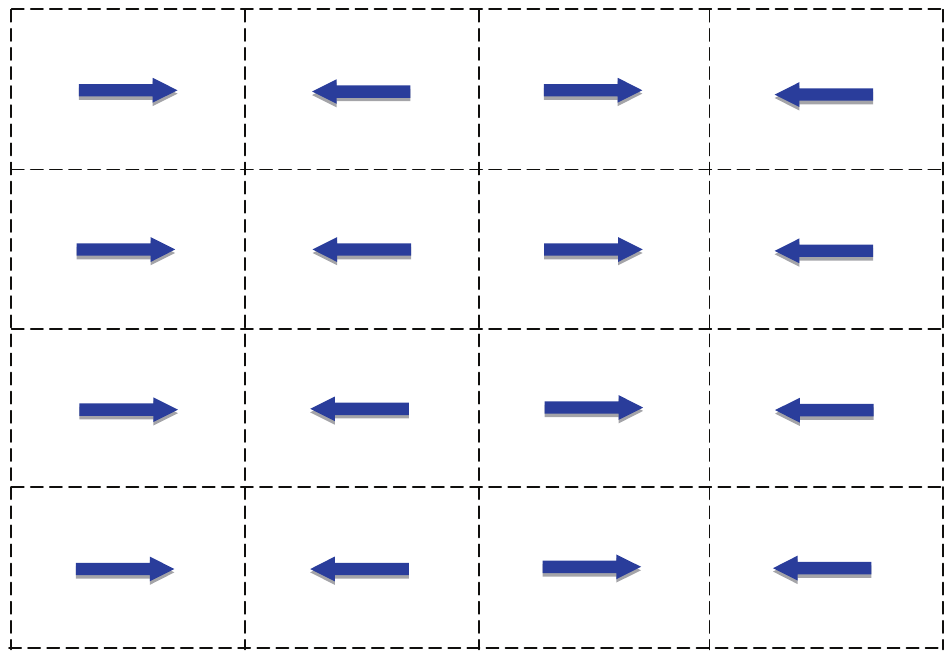}
\caption{Low temperature magneto-structural pattern on the $ab$ plane formed
by the Fe atoms of the iron arsenide systems. The spins order with wavevector
$(\pi, 0)$ forming a collinear antiferromagnet. The associated structural
change is an orthorhombic distortion of the square lattice.}
\label{fig:2}
\end{figure}

Selenium doped iron telluride is represented as Fe$_{1+y}$Te$_{1-x}$Se$_x$ due to the presence
of excess Fe in the interstitial positions of the Te layer.~\cite{gronvold}
The amount of excess Fe decreases with increasing selenium doping, and the phase diagram of the
undoped compound is known to depend upon the amount of excess Fe.~\cite{mizuguchi}
In the following we
focus on the system with $y \approx 0.076$ in the undoped state ($x=0$).
It is known to undergo a first-order transition at around 65 K from a high
temperature paramagnetic tetragonal phase to a low temperature antiferromagnetic monoclinic
phase (throughout the paper we follow the notations of an unit cell with 1Fe/cell).~\cite{bao}
The transition temperature decreases with increasing $x$, but the structural distortion remains
concomitant with the magnetic transition.
This makes the 11 material different from the 1111 and 111 systems, where there are two separate
transitions at all doping, and also from the 122 system where the single transition splits into
two transitions with doping.
More importantly, the symmetry breaking associated with the magneto-structural transition of the
telluride system is different from that of the arsenides. Thus,
the low-temperature phase of the 11 system (Fig.~\ref{fig:1}) is monoclinic and is a
bicollinear antiferromagnet
with wavevector $(\pi/2, \pi/2)$ in the $ab$ plane,~\cite{bao,li,martinelli} in contrast to
the low-temperature phase of the iron arsenides (Fig.~\ref{fig:2})
which is orthorhombic and is a collinear
antiferromagnet with wavevector $(\pi, 0)$.
However, despite these apparent differences,
in the following we show that a feature common to all the iron based superconductors is
the presence of important magneto-elastic couplings whose microscopic origins appear to
be the same.

The magneto-structural transition of Fe$_{1+y}$Te$_{1-x}$Se$_x$ has been studied earlier from a field
theoretic point of view without taking into account all possible magneto-elastic terms.~\cite{xu-hu}
At a more
microscopic level the $(\pi/2, \pi/2)$ magnetic structure is somewhat puzzling from a weakly
correlated band picture because, unlike the case of the iron arsenide materials, the magnetic ordering
does not correspond to a nesting wavevector of the Fermi surface sheets. This has motivated theorists
to seek rationale in models that are based on strong correlation physics such as a localized
spin model~\cite{fang}
or that describing magnetism induced by orbital ordering.~\cite{turner}
Alternately, it has also been argued that
doping due to excess Fe changes the Fermi surface sufficiently such that $(\pi/2, \pi/2)$ is indeed
a nesting wavevector.~\cite{savrasov}

The rest of the paper is organized as follows.
In the next section we describe the magneto-structural transition by a Ginzburg-Landau
free energy whose form is independent of the microscopic details. The $(\pi/2, \pi/2)$ magnetic
structure is described by four $O(3)$ magnetic order parameters which gives rise to several symmetry
allowed magneto-elastic couplings. Thus, the magnetic order parameters are coupled to the monoclinic
component of the uniform strain, as well as to the
lattice distortions associated with wavevectors of high symmetry in the Brillouin zone.
In the magnetic phase the former coupling distorts the lattice monoclinically with long and short
next-nearest-neighbour Fe-Fe bonds (Fig.~\ref{fig:1}a),
while the latter produces long and short
nearest-neighbour Fe-Fe bonds which gives rise to dimerization of the ferromagnetic chains (Fig.~\ref{fig:1}b).
The experimental observation of both these distortions indicate that the
magneto-elastic terms are non-negligible.~\cite{bao,li,martinelli}
Furthermore, we find that the effective magnetic free energy generated by the lattice favours a first
order transition, as observed experimentally. We finish section II by briefly recalling the
mean field description
of the iron arsenides, and by comparing it with the iron telluride system.
In section III we
propose a microscopic hamiltonian which compactly describes the magneto-elastic
properties of all the iron based superconductors, and we conclude in section IV
by pointing out directions for future research.

\section{Ginzburg-Landau theory}

\subsection{Iron telluride}

The Ginzburg-Landau free energy describing the phase transition can be expressed as
\beq
\label{eq:FGL}
F_{GL} = F_M + F_E + F_{ME}
\eeq
corresponding to the magnetic, elastic and the magneto-elastic parts respectively.
Hereafter the notation refers to the $ab$ plane of the system where the symmetry
considerations are non-trivial.

The magnetic sector is described by four $O(3)$ order parameters which
are the Fourier components $\bM_i$ with $i = 1, \ldots, 4,$ of the magnetization ($\bM$)
corresponding to the wavevectors $\bq_1 = (\pi/2, \pi/2) = - \bq_3$, and
$\bq_2 = (-\pi/2, \pi/2) = -\bq_4$. Thus, at the mean field level the magnetization
is given by
$
\bM (\br_n) = \sum_{i=1}^{4} \bM_i \exp (i \bq_i \cdot \br_n),
$
where $\br_n$ denotes a lattice position. Since $\bM (\br_n)$ is real, we have $\bM_1 = \bM_3^{\ast}$
and $\bM_2 = \bM_4^{\ast}$. Alternatively, one can define four real-valued $O(3)$ order parameters
$\bL_i$ with $i = 1, \ldots, 4,$ such that $\bM_1 = [(\bL_1 + \bL_3) - i (\bL_1 - \bL_3)]/2$ and
$\bM_2 = [(\bL_2 + \bL_4) - i (\bL_2 - \bL_4)]/2$. In terms of $\bL_i$ the magnetization is
\[
\bM (\br_n) = \sum_{i=1}^{4} \bL_i \left[ \cos (\bq_i \cdot \br_n) + \sin (\bq_i \cdot \br_n) \right].
\]

The lattice sector is described by means of the strain tensor $\epsilon_{ij}$ which
can be written as $\epsilon_{ij}(\mathbf r_n)=u_{ij} + {i\over 2} \sum_{\mathbf q \neq 0} [q_iu_j(\mathbf q)
+ q_ju_i(\mathbf q) ]e^{i\mathbf q \cdot \mathbf r_n} $. Here $u_{ij}$
describes uniform strains and $\mathbf u(\mathbf q)$ is the Fourier transform of the displacement field
$\bu (\br_n)$.~\cite{LarkinPikin}
At the mean field level and to the lowest order in an expansion of the
free energy in terms of the order parameters,
we find that the elastic variables that couple with the magnetic ones are the
shear component $u_{xy}$ and the quantities $\mathbf u_i \equiv \mathbf u(\mathbf q_i)$
with $i=5,6$ and $7$ corresponding to $\bq_5 = (\pi, \pi)$, $\bq_6 = (\pi, 0)$ and $\bq_7 = (0, \pi)$.

The transformation of the above variables under the symmetry operations of the square
lattice are summarized in Table~\ref{table}, using which one can construct $F_{GL}$
from the terms that are allowed by symmetry.
\begin{table*}[t]
\begin{tabular}{c p{10pt} c p{10pt} c p{10pt} c p{10pt} c p{10pt} c }
\hline \hline
&& && \multicolumn{3}{c}{reflections} && \multicolumn{3}{c}{translations}\\
&& 90$^\circ$ rotation && $x$-axis && $x=y$ axis && along $x$ && along $y$ \\
\hline
${\mathbf L}_1$ && ${\mathbf L}_2$  && ${\mathbf L}_4$ && ${\mathbf L}_1$ && ${\mathbf L}_3$  && ${\mathbf L}_3$   \\
${\mathbf L}_2$ && ${\mathbf L}_3$  && ${\mathbf L}_3$ && ${\mathbf L}_4$ && $-{\mathbf L}_4$ && ${\mathbf L}_4$   \\
${\mathbf L}_3$ && ${\mathbf L}_4$  && ${\mathbf L}_2$ && ${\mathbf L}_3$ && $-{\mathbf L}_1$ && $-{\mathbf L}_1$   \\
${\mathbf L}_4$ && ${\mathbf L}_1$  && ${\mathbf L}_1$ && ${\mathbf L}_2$ && ${\mathbf L}_2$  && $-{\mathbf L}_2$   \\
$u_{xy}$        && $-u_{xy}$        && $-u_{xy}$       && $u_{xy}$        && $u_{xy}$         && $u_{xy}$ \\
$(u_5^x, u_5^y)$ && $(u_5^y, - u_5^x)$ && $(u_5^x, - u_5^y)$ && $(u_5^y, u_5^x)$ && $(-u_5^x, - u_5^y)$ && $(-u_5^x, - u_5^y)$ \\
$(u_6^x, u_6^y)$ && $(u_7^y, -u_7^x)$  && $(u_6^x, -u_6^y)$  && $(u_7^y, u_7^x)$ && $(-u_6^x, -u_6^y)$  && $(u_6^x, u_6^y)$ \\
$(u_7^x, u_7^y)$ && $(u_6^y, -u_6^x)$  && $(u_7^x, - u_7^y)$ && $(u_6^y, u_6^x)$ && $(u_7^x, u_7^y)$    && $(-u_7^x, - u_7^y)$ \\
\hline \hline
\end{tabular}
\caption{Transformation properties of the magnetic and the elastic order parameters under symmetry operations of
the square lattice. The definitions of the various order parameters are given following Eq.~(\ref{eq:FGL}).}
\label{table}\end{table*}
In the following we assume $O(3)$ symmetry in the magnetic sector, which amounts to neglecting
spin-orbit coupling.
The magnetic part of the free energy can be expressed as
\beq
\label{eq:FM}
F_M = \frac{A}{2} \left( \bL_1^2 + \bL_2^2 + \bL_3^2 + \bL_4^2 \right) + \ldots,
\eeq
where $A = A^{\prime} (T - T_N)$ is the only temperature ($T$) dependent coefficient, and $T_N$ is
temperature for the magneto-structural transition. The ellipses denote the fourth order and the
sixth order terms, the latter being necessary because eventually the transition is first order.
At the fourth order there are seven invariant terms namely, (i) $\sum_i L_i^4$, (ii) $L_1^2 L_2^2
+ {\rm cyclic} \ {\rm terms}$, (iii) $(\bL_1 \cdot \bL_2)^2 + {\rm cyclic} \ {\rm terms}$,
(iv) $(\bL_1 \cdot \bL_3)(\bL_2 \cdot \bL_4)$, (v) $(\bL_1 \cdot \bL_2)(\bL_3 \cdot \bL_4)
+ 2 \leftrightarrow 4$, (vi) $(\bL_1 \cdot \bL_3)^2 + (\bL_2 \cdot \bL_4)^2$,
and (vii) $L_1^2 L_3^2 + L_2^2 L_4^2$. Thus, at this order there are seven independent coupling
constants in terms of which the phase diagram described by $F_M$ is rather rich. It is
not the purpose of this paper to investigate how the phase diagram varies with different couplings,
which is anyway a formidable task. The experimentally observed magnetic state is described by
a non-zero value of any one of the $\bL_i$, and the energetics of this choice is clearly beyond
symmetry based arguments and mean field theory. Consequently, we do not write explicitly the terms
beyond quadratic order. Next, the elastic part of the free energy is given by
\beq
\label{eq:FE}
F_E = \frac{c_{66}}{2} u_{xy}^2 + \frac{\Om_1}{2} \bu_5^2 + \frac{\Om_2}{2} \left(
\bu_6^2 + \bu_7^2 \right),
\eeq
where $c_{66}$ is the elastic constant for monoclinic distortion, and $\Om_1$ and $\Om_2$ are
the elastic stiffness of the displacements at the respective wavevectors.
Since the elastic sector is not critical,
it is sufficient to truncate the expansion at the quadratic order. Finally,
using Table~\ref{table}, the lowest order magneto-elastic
part is given by
\begin{align}
\label{eq:FME}
&F_{ME}
=
g_1 u_{xy} \left( L_1^2 + L_3^2 - L_2^2 - L_4^2 \right)
\nonumber \\
&
+ g_2 \left[
u_5^x \left( L_1^2 - L_3^2
-L_2^2 +L_4^2 \right) + u_5^y
\left( L_1^2 - L_3^2 + L_2^2 - L_4^2 \right)
\right]
\nonumber \\
&
+ g_3 \left[ u_6^x \left( \bL_1 \cdot \bL_4 - \bL_2 \cdot \bL_3 \right)
+ u_7^y \left( \bL_1 \cdot \bL_2 - \bL_3 \cdot \bL_4 \right) \right].
\end{align}
In the above we ignore the standard magneto-striction term since it is
present in all materials, and is not peculiar to the 11 system.

We minimize $F_{GL}$ with respect to the elastic degrees of freedom and we get
\begin{subequations}
\begin{align}
\label{eq:uxy}
u_{xy} &= - \frac{g_1}{c_{66}} \left( L_1^2 + L_3^2 - L_2^2 -L_4^2 \right),
\\
\label{eq:u5}
u_5^{x/y} &= - \frac{g_2}{\Om_1} \left( L_1^2 - L_3^2 \mp L_2^2 \pm L_4^2 \right),
\\
\label{eq:u67}
u_6^x &= - \frac{g_3}{\Om_2} \left( \bL_1 \cdot \bL_4 - \bL_2 \cdot \bL_3 \right),
\\
u_7^y &= - \frac{g_3}{\Om_2} \left( \bL_1 \cdot \bL_2 - \bL_3 \cdot \bL_4 \right),
\end{align}\end{subequations}
while the remaining elastic variables are zero at equilibrium. Thus, the lattice
mediated effective magnetic free energy is given by
\begin{align}
\label{eq:FM-prime}
&F_M^{\prime} = \left( F_E + F_{ME} \right)_{\rm equilib}
\nonumber \\
&= - \frac{g_1^2}{2 c_{66}} \left( L_1^2 + L_3^2 - L_2^2 -L_4^2 \right)^2
\nonumber \\
&
- \frac{g_2^2}{2 \Om_1} \left[
\left( L_1^2 - L_3^2
- L_2^2 + L_4^2 \right)^2 + \left( L_1^2 - L_3^2 + L_2^2 - L_4^2 \right)^2 \right]
\nonumber \\
&
- \frac{g_3^2}{2 \Om_2} \left[ \left( \bL_1 \cdot \bL_4 - \bL_2 \cdot \bL_3 \right)^2
+ \left( \bL_1 \cdot \bL_2 - \bL_3 \cdot \bL_4 \right)^2 \right].
\end{align}
The simple exercise above allows us to make the following two points about
the experimentally observed magnetic phase where $L_1 = L$ (say) is the spontaneous
magnetization and the remaining $L_i$ are zero. (i) The lattice undergoes
a monoclinic distortion with $u_{xy} = -g_1 L^2/c_{66}$ producing long and short
next-nearest-neighbour Fe-Fe bonds (Fig.~\ref{fig:1}a).
Simultaneously, the nearest-neighbour Fe-Fe bonds
dimerize such that $u_5^x = u_5^y = - g_2 L^2/\Om_1$ (Fig.~\ref{fig:1}b).
The experimental observation
of both these distortions imply that the magneto-elastic coupling is
non-negligible.~\cite{bao,martinelli}
(ii) The lattice mediated effective magnetic free energy is given by
$F_M^{\prime} = - (g_1^2/c_{66} + g_2^2/\Om_1)L^4/2$, which is a negative contribution
to the free energy at fourth order. As such, this contribution favours the
experimentally observed first order transition.

\subsection{Iron arsenides}

To finish this section we briefly recall the magneto-elastic properties of the
iron arsenides (Fig.~\ref{fig:2}),
studied in detail in Ref.~\onlinecite{cano},
in order to compare them with the 11 material
and to facilitate the discussion in the following section. The relevant magnetic
order parameters for the iron arsenides are the Fourier components $\bM_6$ and
$\bM_7$ of the magnetization corresponding to the wavevectors $\bq_6$ and
$\bq_7$ respectively. They couple to the orthorhombic component $u_{xx} - u_{yy}$
of the uniform strain tensor, and at the mean field level the magneto-elastic part
has the form~\cite{foot1}
\beq
\label{eq:FME-2}
F_{ME}^{\rm FeAs} = g_4 \left( u_{xx} - u_{yy} \right) (M_6^2 - M_7^2).
\eeq
This term determine
(a) if there is a single magneto-structural transition (as in the undoped 122 systems) or two
separate transitions (as in the 1111 or the sufficiently doped 122 systems),
(b) the order (first versus second) of the transitions, and (c) why the system prefers
a collinear magnetic state instead of a non-collinear order.

In the case of the iron
arsenides, since the system is near an orthorhombic transition (this is explicit in the
phase diagram of the 1111 and the doped 122 systems) where the corresponding elastic
constant $c_o$ vanishes, it is possible to argue that the energy scale $g_4^2/|c_o|$
generated magneto-elastically is dominant and controls the main features of the phase
diagram for the magneto-structural transition.
In contrast, the phase diagram of the 11 system does not exhibit a monoclinic
structural transition where $c_{66}$ vanishes, and since the magnitude of $c_{66}$
from ultrasound experiments is currently unavailable to us, a similar argument for the
scale $g_1^2/|c_{66}|$ cannot be made at present. Nevertheless, the distortions of
the lattice of the undoped 122 systems and the 11 systems on entering the magnetic phase
provide clear evidence that the magneto-elastic terms are non-negligible and ubiquitous
for all the iron based superconductors.

\section{Microscopic coupling}
\label{sec:microscopic}

In the previous section we argued in favour of the existence of magneto-elastic couplings in all
the iron based superconductors. The precise form of the couplings at the mean field level
differs between the iron arsenides and the 11 systems because the magnetic order parameters
in these two classes are different. In the following we introduce a microscopic magneto-elastic
hamiltonian whose mean field form captures both Eqs.~(\ref{eq:FME}) and (\ref{eq:FME-2}).
The purpose of such a hamiltonian is two-fold. Firstly, to unify the various magneto-elastic
effects observed in different classes of systems and provide a common description. This is
a first step to understand the microscopic origin of these couplings. Secondly, to go beyond
mean field theory and study the effect of these couplings at the level of fluctuations.

The simplest microscopic magneto-elastic hamiltonian is given by
\beq
\label{eq:ham}
\mathcal{H}_{ME} = \sum_{n,\dl} \lam_{\dl}
\left( {\bf S}_{n + \dl} \cdot {\bf S}_n \right)
\br_{n, \dl} \cdot
\left[ \bu(\br_{n+\dl}) - \bu(\br_n) \right],
\eeq
where $n$ denotes lattice sites, $\dl$ implies nearest-neighbours and next-nearest-neighbours,
$\br_{n,\dl} = \br_{n+\dl} - \br_n$,
$\lam_{\dl}$ is the strength of the coupling that can depend on the bond lengths and the
bond angles, and ${\bf S}_n$ is the electron spin at site $n$. It is easy to verify that
when the spins ${\bf S}_n$ and the displacements $\bu(\br_n)$ are replaced by the appropriate
mean field variables, one obtains the mean field results of Eqs.~(\ref{eq:FME}) and
(\ref{eq:FME-2}). Thus, while the coupling $g_1 = -4 \lambda_{nnn}$ originates
from the next-nearest-neighbour
magneto-elastic interaction $\lambda_{nnn}$;
$g_2$, $g_3$ and $g_4$ are obtained from nearest-neighbour
interaction $\lambda_{nn}$,
implying that in the 11 system, in fact, $g_2 =g_3 = - 4 \lambda_{nn}$.
Experimentally, all
the different kinds of distortions in the 1111, the 122 and the 11 systems
that have been reported are such that the ferromagnetic bonds are
shorter than the corresponding antiferromagnetic bonds,~\cite{cruz,bao,martinelli,private}
which implies that $\lam_{\dl} > 0$.

At present it is being debated whether the iron based superconductors are
weakly interacting metals or those that are strongly interacting. In this
context it should be noted that Hamiltonians of the above type are quite common
in the study of lattice effects in insulating magnetic systems, where
$\lam_{\dl} \propto \ptl J/\ptl r$ is the variation of the Heisenberg exchange coupling
$J$ with distance. Thus, the existence of the magneto-elastic couplings would
seem to favour the point of view that the
magnetic properties of these systems are more suitably described by localized spin
models rather than a weak coupling band picture.~\cite{qimiao}
On the other hand, it is also possible to interpret the above Hamiltonian from a
band picture with moderately strong interaction. Within a semi-classical paramagnon theory
of magnetism, which is suitable for band metals, ${\bf S}_n$ can be taken as a paramagnon
field which is an $O(3)$ variable describing local fluctuations of magnetization.
Therefore, while it is clear that this  is an effect of
electron-lattice coupling in an environment where
electron-electron interaction is crucial, it is difficult at this point to conclude
whether a localized limit is necessary for describing such interaction.

\section{Summary}

To summarize, in this paper we studied the effects of magneto-elastic couplings on the
magnetic phase transition in Fe$_{1+y}$Te$_{1-x}$Se$_x$.
From symmetry considerations we showed that on entering the magnetic phase,
these couplings give rise to
uniform monoclinic distortion (Fig.~\ref{fig:1}a) of the unit cell with one Fe/cell,
and also induce
dimerization (Fig.~\ref{fig:1}b) of the ferromagnetic Fe chains
due to alternate lengthening and shortening of the nearest-neighbour Fe-Fe bonds.
We also showed that the effective
magnetic energy generated by the lattice favours a first order transition. Finally,
we compared this system with the iron arsenide systems, and we proposed a microscopic
Hamiltonian to describe the magneto-elastic effects in both these classes of iron based
superconductors.
Microscopically, this is a manifestation of electron-lattice coupling in a system where
electron-electron interaction is strong.
In the future we hope to study the effect of these couplings on the
magnetism and the superconductivity.

\begin{acknowledgments}

We are very thankful to W. Bao, O. Cepas, M. Civelli A. Martinelli, A. Vishwanath, and T. Ziman
for insightful discussions. I.P. is thankful for the hospitality of the
Indian Association for the Cultivation of Sciences where this work was initiated.
K.S. thanks DST, India for support through project no. SR/S2/CMP-001/2009.
\end{acknowledgments}


\begin{thebibliography}{99}
\bibitem{foot1} Here the notations are different from Ref.~\onlinecite{cano} because the latter
adopted the convention for a unit cell with two Fe/cell. Expressed in the notations of this paper
the magnetic order parameters in Ref.~\onlinecite{cano} are $\bL_{1/2} = \bM_6 \pm \bM_7$, and
the monoclinic distortion in Ref.~\onlinecite{cano} is equivalent to an orthorhombic distortion
when the principal axes are rotated $45^{\circ}$.
\end{thebibliography}
\end{document}